\documentclass[conference]{IEEEtran}
\usepackage{cite}
\usepackage{amsmath,amssymb,amsfonts, bm}
\usepackage{algorithmic}
\usepackage{algorithm,float}
\usepackage{graphicx, subfigure}
\usepackage{textcomp}
\usepackage{xcolor}
\usepackage{multirow}
\usepackage{balance}
\usepackage{colortbl}
\usepackage{url}
\usepackage{cuted}
\def\BibTeX{{\rm B\kern-.05em{\sc i\kern-.025em b}\kern-.08em
		T\kern-.1667em\lower.7ex\hbox{E}\kern-.125emX}}
\newcommand\Tstrut{\rule{0pt}{2.6ex}}	
	
\begin{document}
	\title{Complex Rotation-based Linear Precoding for Physical Layer Multicasting and SWIPT}
	
	\author{%
		\IEEEauthorblockN{Xinliang Zhang and Mojtaba Vaezi}
		\IEEEauthorblockA{
			Department of Electrical and Computer Engineering,\\
			Villanova University, Villanova, PA 19085, USA\\
			Emails:  \{xzhang4, mvaezi\}@villanova.edu
		}
	}
	\maketitle
	\begin{abstract}
		With the goal of improving spectral efficiency, complex rotation-based precoding and power 
		allocation schemes are developed for two  multiple-input  multiple-output 
		(MIMO) communication systems, namely, simultaneous wireless 
		information and power 	transfer (SWIPT) and 
		physical layer multicasting.
		While the state-of-the-art solutions for these problems use very different approaches, the proposed approach treats them similarly using a general tool and works efficiently 		for any number of antennas at each node. 
		 Through modeling the 
		precoder using complex rotation matrices,   
		objective functions (transmission rates) of the above systems can be formulated and solved 
		in a similar structure.    
		Hence, this approach simplifies signaling design for MIMO systems and can reduce the hardware complexity by having one set of parameters to optimize.
		Extensive numerical results show that the proposed  approach  
		 outperforms state-of-the-art  
		solutions for both problems. It increases transmission rates for  multicasting and achieves higher rate-energy 
		regions in the SWIPT case. In both cases, the improvement is significant 
		($20\%$-$30\%$) in practically important settings where the users have 
		one or two antennas. Furthermore, the new precoders are less time-consuming than the existing solutions.   
	\end{abstract}
	\section{Introduction}\label{sec_intro}
   

{Efficient  spectrum  usage and management are of vital  importance in modern communication systems. To meet the communication requirements of 
	multiple users, 
	efficient 
	\textit{spectrum management} and  \textit{spectrum sharing} are
	required to  capture 
	and schedule
	the best available spectrum  \cite{akyildiz2006next}. 
	By exploiting multipath propagation, multiple-input multiple-output (MIMO)
	communication has significantly improved the spectrum efficiency and has become an essential element of today's wireless communication standards.
	It is known that the design  of the precoder at the transmitter largely affects
	the performance of MIMO	systems. 
From an information-theoretic perspective,  \textit{linear precoding} is optimal to achieve  high spectral efficiency \cite{vu2007mimo}; it also  keeps  the complexity low
    \cite{skoglund2003capacity}.} 


Besides spectral efficiency and high-speed transmission, in modern wireless systems, there is a desire to improve  multiple other objectives, such as energy-efficient,  multicasting, and security. 
Such needs have motivated linear precoding for various objectives such as wireless information transfer 
(WIT), 
energy harvesting (EH) \cite{zhang2013mimo}, simultaneous wireless 
information and 
power transmission (SWIPT) \cite{zhang2013mimo, rostampoor2017energy}, physical layer
multicasting \cite{sidiropoulos2006transmit, zhu2012precoder},  and so forth. 
While some of the above  problems have analytical precoding design--for 
example,  singular value decomposition (SVD) for WIT 
\cite{cover2012elements} and EH \cite{zhang2013mimo}--some others such as
SWIPT and multicasting transmission only have numerical optimization based solutions.
Moreover, these solutions differ from a problem to another since their optimization problems have different structures.

Actually,  all of those problems can be solved using one general
approach, named rotation-based precoding.  This approach simplifies   
positive semidefinite (PSD) constraint to a set of linear constraints and 
enables us to solve the related problems using general optimization tools like 
\texttt{fmincon} in MATLAB. 
The rotation-based precoding was first proposed in \cite{vaezi2017journal} for the MIMO wiretap channels with two antennas.  Later, in \cite{zhang2019rotation}, this method was extended  to the cases for an arbitrary number of antennas at each node. This method applies Givens rotation to build the precoding matrix to maximize secure transmission rates.
In \cite{tavangaran2020mimo}, this method was applied to energy harvesting maximization with a secrecy rate constraint.  In all of these cases, there is merit in applying this method to design precoder matrices. Rotation-based precoding can provide  an analytical solution for the cases in which the number of transmit antennas is two \cite{vaezi2017journal, tavangaran2020mimo}.   For other cases, this method is shown to provide more robust solutions than existing solutions  \cite{vaezi2019rotation, zhang2019rotation}. Overall, the rotation-based  method is a general tool for precoding design and has great potential for extension to other related problems \cite{zhang2019rotation}. The above solutions are, however,  based on real rotation matrices which work only for real-valued channels. 

In this paper, we develop complex rotation-based precoders for two 
applications: SWIPT and physical layer multicasting.  	
In SWIPT,  one user receives information 
at a high rate, and the other user simultaneously harvests energy from the common transmitter.  Linear precoding and power allocation solution can be 
iteratively optimized in \cite{zhang2013mimo} using time-switching and power-splitting (TS-PS). 
Multicasting is used when a transmitter broadcasts  public messages, such as advertisements  and forecasts. This is a min-max fair problem to enlarge the transmission rate for all users. In the MIMO case, a cyclic alternating ascent (CAA) precoder  is proposed in \cite{zhu2012precoder}. In the multiple-input single-output (MISO) case, semidefinite relaxation (SDR) techniques yield a closed-form solution \cite{sidiropoulos2006transmit}.
These methods yield  approximate solutions and their performance  is 
affected by the number of antennas. 

	 
	The rotation-based precoders we develop	for the above-mentioned two problems improve the performance, work for an arbitrary number of antennas at each node, reduce the computational complexity, and can be applied to complex- and real-valued channels. The 
	main contributions of this paper  are 
	summarized as follows:
	\begin{itemize}
		\item For \textit{SWIPT}, our precoder is more robust than state-of-the-art  solutions and increases the data transmission rates particularly when harvesting users have one or two antennas (which is of great importance in practice).  In particular,  the rate improvement can reach as high as 20\% when compared with the TS-PS \cite{zhang2013mimo}.
		\item For \textit{multicasting}, our precoder enlarges the data transmission rates  
		while  lowering the computational complexity compared to the state-of-the-art methods. For example, compared to the SDR, the rate improvement can be up to nearly 30\% for practical antenna settings while reducing the computational complexity more than an order of magnitude. 				
	\end{itemize}

	The  remainder of this paper is organized as follows. Channel 
	models and objective functions are discussed in Section~\ref{sec_sys}. The 
	 precoder design is discussed in Section~\ref{sec_rot}. Numerical
	results are illustrated in Section~\ref{sec_data}. Finally, we conclude the 
	paper in Section~\ref{sec_conclu}.
	

	\begin{table*}[t]
		\caption{The Desired  Objectives and Existing Solutions}
		\label{tab_modes}
		\centering
		\begin{tabular}{l|p{6cm}|p{6cm}}
		 \hline 
		\Tstrut	\textbf{Problems} 
			& \textbf{SWIPT} 
			& \textbf{Multicasting} 	\\  \hline\hline
			\Tstrut
			Objectives
			& \scalebox{1.1}{$\begin{array}{l}
			\Tstrut	\mathcal{C}_1 \triangleq
				\max\limits_{\mathbf{Q}} \log_2 
				{|\mathbf{I}_{n_1}
					+\mathbf{H}_1\mathbf{Q}\mathbf{H}_1^H|}\\
				\qquad{\quad\rm s.t.\;}{\rm tr}
				(\mathbf{H}_2\mathbf{Q}\mathbf{H}_2^H)\geq {\cal \bar E}
				\end{array}$}
			&\scalebox{1.1}{$\begin{array} {l}
				\mathcal{C}_2 \triangleq
				\max\limits_{\mathbf{Q}}
				\; \min\limits_{u=1,2}
				\log_2{|\mathbf{I}_{n_u}+\mathbf{H}_u\mathbf{Q}\mathbf{H}_u^H|}
				\end{array}	$}
			\\ \hline
			\Tstrut
			Reference 
			& TS-PS\cite{zhang2013mimo}
			& SDP\cite{sidiropoulos2006transmit}, CAA\cite{zhu2012precoder}\\
			\hline
		\Tstrut	Function
			& Balances the transmission rate to Rx1 as well as energy harvested at 
			Rx2.
			& Transmits the same information to Rx1 and Rx2. The maximum speed 
			is restricted by the low-speed user. 
			\\ \hline
			
		\end{tabular}
	\end{table*}

	\section{Channel Models and  Preliminaries}\label{sec_sys}
	\subsection{Channel Models}
	Consider a two-user MIMO wireless communication system 
	with one transmitter (Tx) and two receivers as shown in 
	Fig.~\ref{fig_figTxRx}. The Tx is equipped with   
	$m$ antennas and sends information to two receivers. At the Tx, 
	a linear precoder is applied, in which 
	$\mathbf{s}\triangleq[s_1,\hdots,s_{m}]$  denotes an independent and unit 
	power symbol vector, that is, 
	$\mathbb{E}\{\mathbf{s}\mathbf{s}^H\}=\mathbf{I}_m$.   
	$\mathbf{\Lambda}\triangleq{\rm diag}(\lambda_1,\hdots,\lambda_{m})$ 
	represents the power allocation factors and    
	$\mathbf{V}\in\mathbb{C}^{m\times m}$ is the precoding matrix,  
	$\mathbf{\Lambda}$ and $\mathbf{V}$  together define  precoding and power allocation, where 
	the transmitted signal is 
	\begin{align}
	\mathbf{x} =\mathbf{V}\mathbf{\Lambda}^{\frac{1}{2}}\mathbf{s}.
	\end{align}
	Then, the covariance matrix of $\mathbf{x}$ 
	becomes 
	$\mathbf{Q}\triangleq\mathbb{E}\{\mathbf{x}\mathbf{x}^H\}
	=\mathbf{V}\mathbf{\Lambda}\mathbf{V}^H$. At the receiver side, 
	receiver~1 (Rx1)
	and receiver~2 (Rx2) are equipped with  $n_1$ 	and $n_2$  antennas. They
	access and decode the information for their users respectively. Assuming 
	transmission is over flat fading channels,  the received signals can be 
	formed 	as  
	\begin{subequations}
	\begin{align}
	\mathbf{y}_1 = \mathbf{H}_1\mathbf{x} + \mathbf{w}_1,\\
	\mathbf{y}_2 = \mathbf{H}_2\mathbf{x} + \mathbf{w}_2,\label{eq_recSig}
	\end{align}
	\end{subequations}
	in which  ${\mathbf{H}_1} \in 
	\mathbb{C}^{n_1 \times m}$  and $\mathbf{H}_2 
	\in \mathbb{C}^{n_2 \times m}$ are the channel matrices corresponding to  
	Rx1 and Rx2,  and $\mathbf{w}_1 \in \mathbb{C}^{n_1\times 1}$ 
	and $\mathbf{w}_2\in \mathbb{C}^{n_2\times 1}$ are independent and 
	identically distributed (i.i.d.) Gaussian noises 
	with zero means and unit variances. 

	\subsection{Objective Functions and State-of-the-Art Solutions}
\begin{figure}[t]
		\centering
		\includegraphics[width=0.48\textwidth]{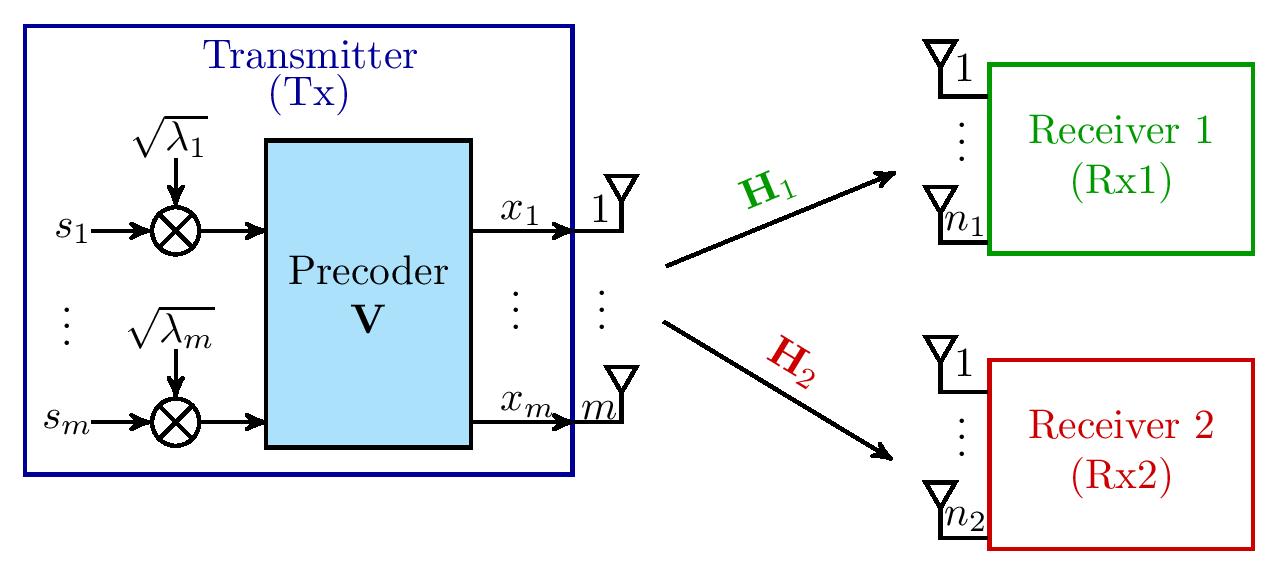}
		\caption{A MIMO system with a transmitter and two 
			receivers. The transmitter applies a linear precoder. The number of 
			antennas at each node ($m, n_1, n_2$)  is arbitrary. }
		\label{fig_figTxRx} 
	\end{figure}

	Many wireless communication systems aim to build a green and  
	secure ecosystem with a variety of services, including 
	SWIPT and  multicasting. 
{Signaling should be designed in a way that the two  users  in Fig.~\ref{fig_figTxRx}  efficiently access  the 
spectrum resource.}
	Essentially,  these problems require input covariance matrix design,  or 
	equivalently precoding and power allocation matrices,  that can maximize 
	the corresponding 
	objective functions. The goal is to find analytical solutions or effective 
	numerical approaches for each problem. 
	In the following, the objective functions of these 
	communications services are explained. 
	\subsubsection{SWIPT}\label{sec_sys_t3-5}
	When Rx1 requires WIT and Rx2 orders EH at the same time, the SWIPT problem arises which needs a balanced 
	precoder that can ensure the  wireless transmission from Tx to 
	Rx1 and energy harvesting from Tx to Rx2  simultaneously and efficiently. 
	As defined in 
	\cite{zhang2013mimo}, SWIPT characterizes the optimal trade-off between 
	the maximum energy and information transfer by the rate-energy boundary, 
	which is given by 
	\begin{subequations}
		\begin{align}\label{eq_opt3} 
		\textmd{(P1)}
		\quad\mathcal{C}_1 \triangleq& 
		\max\limits_{\mathbf{Q}}\cdot\log_2 
		{|\mathbf{I}_{n_1}
			+\mathbf{H}_1\mathbf{Q}\mathbf{H}_1^H|},\\
		&{\;\;\rm s.t.\;} \ 
		\eta\cdot{\rm tr}
		(\mathbf{H}_2\mathbf{Q}\mathbf{H}_2^H)\geq {\cal \bar E}, \\
		&\; \qquad \mathbf{Q}\succeq\mathbf{0}, \mathbf{Q}=\mathbf{Q}^H,
		{\rm tr}(\mathbf{Q})\leq P, \label{eq_3c}
		\end{align}
	\end{subequations}
	in which $\eta$ is the  \textit{converting rate} of  the harvested energy 
	(we assume $\eta=1$ throughout the paper), 
	$\cal\bar{E}$ is a threshold 
	representing the required EH level  at Rx2, and  $P$ is the 
	transmit power. The value of $\cal\bar{E}$ is 
	between the minimum (${\cal 
		E}_{\min}$) and maximum (${\cal 
		E}_{\max}$) harvested energy level, that is
	\begin{align}\label{eq_thres}
	{\cal{\bar{E}}}\triangleq{\cal E}_{\min} + q ({\cal E}_{\max}-{\cal E}_{\min}),
	\end{align}
	where $q$ is a factor between $0\%$ and $100\%$. Since we are looking 
	for the maximum 
	rate-energy boundary, ${\cal E}_{\min}$ is defined as the energy received  
	by Rx2 when 
	Rx1 reaches its optimal data rate. This can be obtained by solving
	\begin{align}\label{eq_opt1}
	\textmd{(P1-1)}
	\max\limits_{\mathbf{Q}\succeq\mathbf{0}, {\rm tr}(\mathbf{Q})\leq 
		P}\log_2 
	{|\mathbf{I}_{n_1}+\mathbf{H}_1\mathbf{Q}\mathbf{H}_1^H|},
	\end{align}
	which is the well-known WIT problem with an analytical solution 
	\cite{cover2012elements} $\mathbf{Q}_1^\ast$ 
	by using singular value decomposition (SVD) and water filling (WF). Then, 
	we 
	have ${\cal E}_{\min}=\eta\cdot{\rm 
		tr}(\mathbf{H}_2\mathbf{Q}_1^\ast\mathbf{H}_2^H)$. On the contrary, 
		${\cal 
		E}_{\max}$ can be obtained when 
	Rx2 reaches the maximum EH level, i.e.,
	\begin{align}\label{eq_opt2}
	\textmd{(P1-2)}
	\quad{\cal 
		E}_{\max} \triangleq& \max\limits_{\mathbf{Q}\succeq\mathbf{0}, {\rm 
			tr}(\mathbf{Q})\leq 
		P}\eta\cdot{\rm 
		tr}(\mathbf{H}_2\mathbf{Q}\mathbf{H}_2^H),
	\end{align}
	which is a typical EH problem and the solution is given in 
	\cite{zhang2013mimo} in an 
	analytical form.
	So, when 
	$q=0\%$, (P1)   degenerates to (P1-1); When $q=100\%$, (P1) is equivalent 
	to (P1-2); 
	Otherwise, $q$ controls the balance between the WIT for Rx1 and 
	EH for Rx2. Since both (P1-1) and (P1-2) are related to the channels, ${\cal\bar E}$ is a dynamic threshold that varies from 
	$\mathbf{H}_1$ and $\mathbf{H}_2$ and is controlled by $q$. 
	Although numerical solutions are studied in \cite{zhang2013mimo, 
		rostampoor2017energy} for SWIPT, these  
	solutions are specifically designed  and they are difficult to be generalized 
	for 
	other precoding problems as a unified solution.

	\subsubsection{Multicasting}\label{sec_sys_t7}
	The Tx broadcasts a public message, such as advertisements and weather 
	forecasts, to all users.
	In this 
	case, the performance of the multicasting rate is subject to the
	minimum rate of the receivers. Therefore, this problem is to maximize the 
	rate of the receiver with the minimum  
	WIT rate in the context of  common 
	information  
	\cite{sidiropoulos2006transmit}. The capacity region of the Gaussian MIMO 
	multicasting is formulated as 
	\begin{subequations}
		\begin{align}\label{eq_opt5}
		\textmd{(P2)}
		\quad\mathcal{C}_2\triangleq& \max\limits_{\mathbf{Q}}
		\; \min\limits_{k=1,2}
		\log_2{|\mathbf{I}_{n_k}+\mathbf{H}_k\mathbf{Q}\mathbf{H}_k^H|}
		\\
		&{\;\;\rm s.t.\;}\mathbf{Q}\succeq\mathbf{0},\mathbf{Q}=\mathbf{Q}^H, 
		{\rm tr}(\mathbf{Q})\leq P. 
		\end{align}
	\end{subequations}
	In the MISO case,  \textit{semidefinite relaxation} 
	(SDR) techniques yield a closed-form solution. In the MIMO case, linear 
	precoding methods are 
	popular for their lower complexity. A \textit{cyclic alternating ascent} (CAA) is 
	proposed in 
	\cite{zhu2012precoder}, and \cite{tan2015opportunistic} mentions that the 
	problem can be 
	solved by \textit{semidefinite programming} (SDP)  solver directly.
	Moreover, a real-valued rotation-based 
	precoding 
	method together with random search was proposed in 
	\cite{vaezi2019rotation} for integrated services with common and 
	confidential 
	messages.  However, the existing methods are limited by the computational 
	complexity and are only available in some cases of antennas.

	In summary, previous works  resort to  
	different solutions to solve for  (P1)-(P2). 
	We propose a unified solution for (P1)-(P2), and the proposed approach is 
	robust to the changes in the number of antennas at the Tx, Rx1, and Rx2.

	\section{The Rotation Modeling as a General Precoding Tool}\label{sec_rot}
	The aforementioned precoding problems are a series of optimization 
	problems on the covariance matrix $ \mathbf{Q} $. The PSD and 
	symmetry requirements  of  $\mathbf{Q}$ increase 
	the difficulty of reaching the optimal solution. Our proposed  
	\textit{complex rotation modeling} (CRM)  circumvents these constraints
	by converting them to a set of linear constraints, as explained in the following.
	
	\subsection{Complex Rotation Modeling (CRM)}
	The covariance matrix $\mathbf{Q}$  can be formed using 
	the eigenvalue decomposition as $\mathbf{Q}\triangleq \mathbf{V} 
	\mathbf{\Lambda} \mathbf{V}^H$.
	$\mathbf{\Lambda}\in\mathbb{C}^{m\times m}$ is a diagonal matrix, 
	whose 
	diagonal elements $[\lambda_1,\hdots,\lambda_{m}]$ are 
	non-negative due to the PSD constraint.
	Moreover, the average power constraints in (P1)-(P2) are equivalent to 
	$\sum_{i=1}^{m}\lambda_i\leq P.$
	To this end, the PSD and power constraints can be represented as 
	a set of  linear constraints
	\begin{align}\label{eq_allcon1}
	\{\lambda_{i}| \lambda_{i}\geq0,  \sum_{i=1}^{m}\lambda_i\leq P\}.
	\end{align}
	Besides, $ \mathbf{V}\in\mathbb{C}^{m\times m} $ is a unitary 
	matrix 
	due to the 
	symmetric property. Then it can be modeled by 
	complex Given's matrix \cite{matrix}  or named rotation matrix as
	\begin{align}\label{eq_Vnbyn_}
	\mathbf{V}=\prod_{i=1}^{m-1}\prod_{j=i+1}^{m} \mathbf{V}_{i,j},
	\end{align}
	where $ \mathbf{V}_{i,j} $ is equal to an identity matrix 
	$\mathbf{I}\in\mathbb{C}^{m\times m}$ except for the following four 
	elements
	\begin{align}\label{eq_VnDsub}
	\left[
	\begin{matrix} 
	v_{i,i}	&v_{i,j}\\
	v_{j,i}	&v_{j,j}
	\end{matrix}
	\right]
	=\left[
	\begin{matrix}
	\cos\theta_{i,j}	&-e^{-j\phi_{i,j}}\sin \theta_{i,j}\\
	e^{j\phi_{i,j}}\sin\theta_{i,j}	&\cos \theta_{i,j}
	\end{matrix}
	\right],
	\end{align}
where $v_{a,b}$, $a,b\in\{i,j\}$, represents the entry $(a,b)$ 
	of 
	matrix 
	$\mathbf{V}_{i,j}$.
	Totally, it requires $n_\alpha=m(m-1)$
	rotation angles to represent $\mathbf{V}$ in \eqref{eq_Vnbyn_}. There is 
	no  constraint on rotation angles, i.e., $\theta_{i,j},\phi_{i,j}\in\mathbb{R}$. 
	Therefore, the optimization on $\mathbf{Q}$ can 
	be equivalently transferred to optimize the parameters using CRM
	with a linear constraint as \eqref{eq_allcon1}.

	\addtolength{\topmargin}{0.01in}
	It is worth mentioning that the order of multiplication in \eqref{eq_Vnbyn_} 
	is 
	not 
	unique and different orders will lead to different rotation angles 
	$\theta_{i,j}$. 
	In this paper, without loss of generality,\footnote{If the order in \eqref{eq_Vnbyn_} is changed, $\mathbf{V}$ will be a different matrix. However, $\mathbf{Q}$ can be maintained the same if the order of diagonal elements of $\mathbf{\Lambda}$ is changed accordingly. A detailed analysis is given in \cite{zhang2019rotation}.} we use the order definition in 
	\eqref{eq_Vnbyn_}. Then, the rotation parameter vector can be defined as
	\begin{align}\label{eq_allcon2}
	\mathbf{r}\triangleq[{\bm\lambda},{\bm\theta},{\bm\phi}]^H,
	\end{align}
	where
	\begin{subequations}
		\begin{align}
		{\bm\lambda}&\triangleq[\lambda_{1},\hdots,\lambda_{m}],\\
		{\bm\theta}&\triangleq[\theta_{1,2},\hdots,\theta_{m-1,m}],\\
		{\bm\phi}&\triangleq[\phi_{1,2},\hdots,\phi_{m-1,m}].
		\end{align}
	\end{subequations}
	To this end, $\mathbf{Q}$ can be identified as a specific covariance matrix 
	by the given 
	parameter vector $\mathbf{r}$, containing both $\bm\lambda$ and 
	$\bm\theta$. The 
	constraint becomes $\mathbf{A}\mathbf{r}\leq\mathbf{b}$, where
	\begin{align}
	\mathbf{A}\triangleq\left[
	\begin{matrix}
	-\mathbf{I}_m & \mathbf{0}_{m\times n_\alpha}\\
	\mathbf{1}_{1\times m} & \mathbf{0}_{1\times n_\alpha}
	\end{matrix} 
	\right],\;
	\mathbf{b}\triangleq\left[
	\begin{matrix} 
	\mathbf{0}_{1\times m}\\
	P
	\end{matrix}
	\right], 
	\end{align}
	in which  $\mathbf{0}_{a\times b}$ (or  $\mathbf{1}_{a\times b}$) is 
	a matrix with all zeros (or ones).
	
	\subsection{CRM for Precoding}
	 The fact that CRM converts the PSD constraint  to a set of
	linear constraints in each problem, facilitates reshaping (P1)-(P2) to new 
	optimization problems (see 
	(P1a)-(P2a) in the following) and use general optimization 
	tools to solve them. In this manner, we do not need to find the solution for each problem via a different approach.
	\subsubsection{CRM for SWIPT}\label{sec_rot_t3-5}
	Applying the CRM to (P1), the objective function of  
	SWIPT becomes
	\begin{subequations}
		\begin{align}\label{eq_opt3a} 
		\textmd{(P1a)}
		\quad\mathcal{C}_1 \triangleq
		& \max\limits_{\mathbf{r}}\log_2 
		{|\mathbf{I}_{n_1}
			+\mathbf{H}_1\mathbf{Q}\mathbf{H}_1^H|},\\
		&{\;\;\rm s.t.\;}\mathbf{A}\mathbf{r}\leq\mathbf{b},\label{eq_P3a_c1}\\
		&\qquad\ 
		\eta\cdot{\rm tr}
		(\mathbf{H}_2\mathbf{Q}\mathbf{H}_2^H)\geq {\cal \bar 
			E},\label{eq_P3a_c2}
		\end{align}
	\end{subequations}
	where $\mathbf{Q}$ is a function of $\mathbf{r}$. 
	This problem can be solved by a general optimization tool such as 
	\texttt{fmincon} in 
	MATLAB. \eqref{eq_P3a_c1} is a linear inequality 
	constraint  and \eqref{eq_P3a_c2} can 
	be treated as a non-linear constrain in \texttt{fmincon}. The initialization of 
	$\mathbf{Q}$ is given by the solution of (P1-2), so that 
	\eqref{eq_P3a_c1} and \eqref{eq_P3a_c2} can be guaranteed. Then, we get 
	the  
	initial
	value of $\mathbf{r}$  using Algorithm~1 in 
	\cite{zhang2019rotation}.

	\subsubsection{CRM for Multicasting} \label{sec_rot_t7}
	(P2) can be reformed as
	\begin{subequations}
		\begin{align}\label{eq_opt5a}
		\textmd{(P2a)}
		\quad\mathcal{C}_2 =& \max\limits_{\mathbf{r}}
		\; \min\left\{
		R_1, R_2
		\right\},\\
		&{\;\;\rm s.t.\;}\mathbf{A}\mathbf{r}\leq\mathbf{b},\label{eq_p5_cons}
		\end{align}
	\end{subequations}
	where $R_1$ and $R_2$ are WIT rates of Rx1 and Rx2, i.e., 
	\begin{subequations}
		\begin{align}
		R_k \triangleq 
		\log_2{|\mathbf{I}_{n_k}+\mathbf{H}_k\mathbf{Q}\mathbf{H}_k^H|},
		\; k=1,2.
		\end{align}
	\end{subequations}
	Then the objective function can be considered as the minimum of two 
	(P1-1) 	with different channels. Since (P1-1) is concave 
	\cite{cover2012elements}, 	(P2a) 	is 
	also concave. It can be solved by considering the following three sub-cases:
	\begin{enumerate}
		\item $R_1\leq R_2$, when $R_1$ has reached its optimal. \\
		That is, solve $\mathbf{Q}_1^\ast$ as the solution of
		(P1-1) and the achievable rate of Rx1 is defined as $R_1^\ast$.
		If $R_1^\ast\leq 
		R_2(\mathbf{Q}_1^\ast)$
		%
		is satisfied, $\mathbf{Q}_1^\ast$ is the solution of (P2a).

		\item $R_2\leq R_1$, when $R_2$ has reached its optimal.\\
		This is the case that by obtaining $\mathbf{Q}_2^\ast$ 
		as 	the solution of 
		\begin{align}\label{eq_opt7b}
		R_2^\ast=
		& \max\limits_{{\rm tr}(\mathbf{Q}_2)\leq P}\log_2 
		{|\mathbf{I}_{n_2}+\mathbf{H}_2\mathbf{Q}_2\mathbf{H}_2^H|},
		\end{align}
		which also has an analytical solution as (P1-1). If $R_2^\ast\leq 
		R_1(\mathbf{Q}_2^\ast)$
		%
		is satisfied,  $\mathbf{Q}_2^\ast$ is the solution of (P2a).
		\item Otherwise,  the solution of (P2a) can be obtained using 
		\texttt{fmincon} with linear inequality constraint as \eqref{eq_p5_cons}. 
		The initialization of rotation 	parameters can be obtained from 
		$\mathbf{Q}_1^\ast$ or $\mathbf{Q}_2^\ast$.
	\end{enumerate}
	Since the first two sub-cases are a WIT problem with analytical 
	solutions, the efficiency of the solution has been improved.
	

\subsection{CRM for More than Two Users}
The CRM-based solution 
can be applied to cases with multiple users.  
With more than two users,  the 
rotation parameters may increase or remain the same depending on the 
specific problem. For example, for multicasting, there is always one 
covariance matrix to be optimized even if the number of users is more than 
two    \cite{zhu2012precoder}, {in which CRM will be efficient in 
this case}. 
On the other hand, if the number of covariance matrices increases by introducing more 
users in the network,  CRM should be applied for each covariance matrix. The capacity region of SWIPT with multiple users is still an open problem. We take  MIMO transmission with multiple users \cite{weingarten2004capacity} and Gaussian 
MIMO multi-receiver wiretap channel \cite{ekrem2011secrecy} as examples, the covariance matrix 
$\mathbf{Q}$ is 
constructed by $U$ independent covariance matrices, i.e.,
$\mathbf{Q} = \sum_{u=1}^{U}\mathbf{Q}_u$. Applying CRM 
to user $u$, 
we have $\mathbf{Q}_u = \mathbf{V}_u \mathbf{\Lambda}_u 
\mathbf{V}_u^H$. Thus, we will have independent  rotation parameters for 
each user to optimize.

\section{Simulation Results}\label{sec_data}
In this section, we evaluate the performance of  CRM  in  SWIPT
and multicasting, respectively.  
In general, CRM can reach the SWIPT rate-energy region. 
For multicasting, CRM can achieve a higher transmission rate with better 
efficiency.
\subsection{SWIPT}\label{sec_data_SWIPT}
\begin{figure}[t]
	\centering
	%
	\subfigure[Corresponding to the channels in 
	\eqref{eq_ch2}.]{\includegraphics[width=0.45\textwidth]{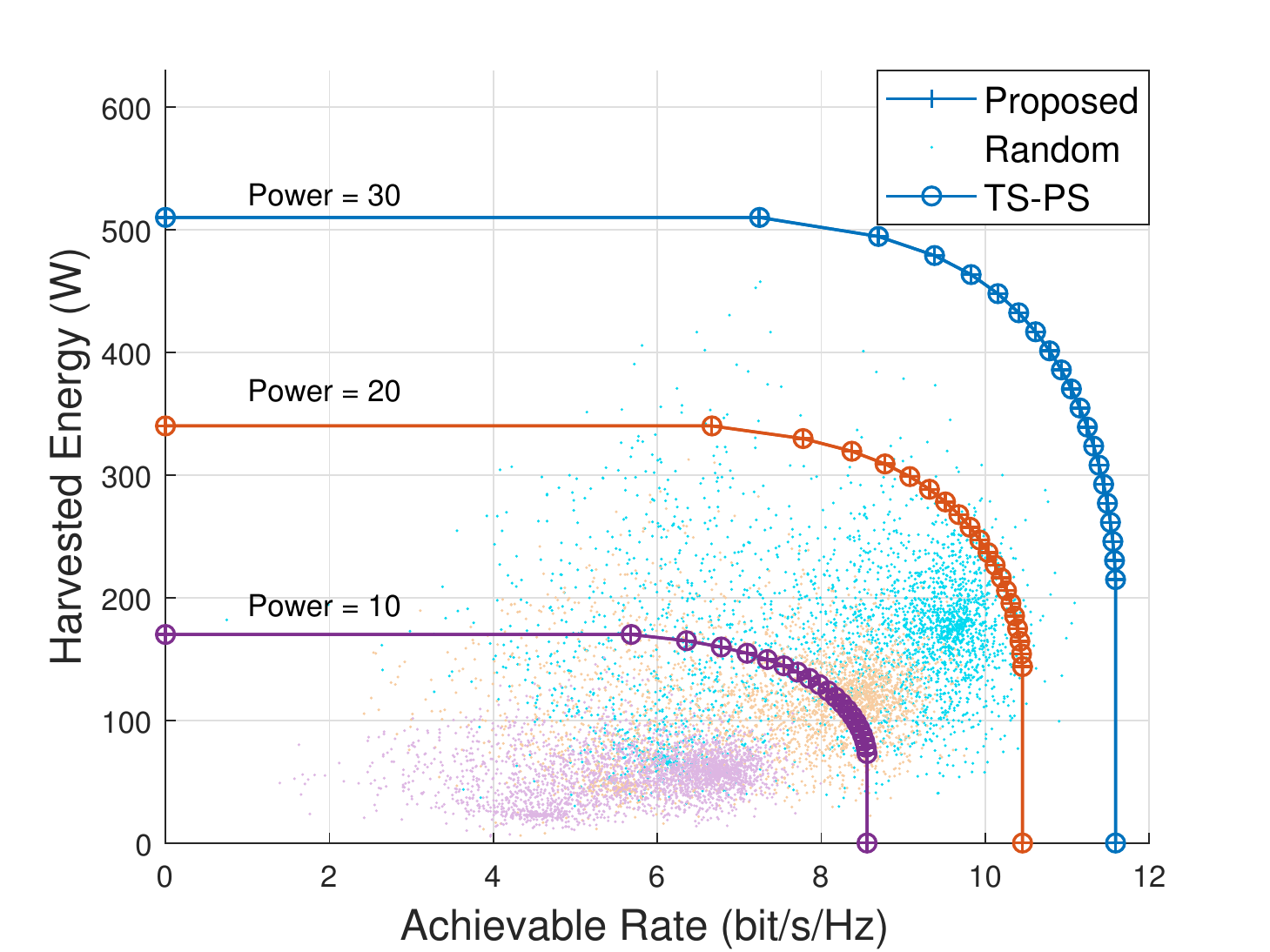}
		\label{fig_SWIPTcurve2}}
	\subfigure[Corresponding to the channels in 
	\eqref{eq_ch3}.]{\includegraphics[width=0.45\textwidth]{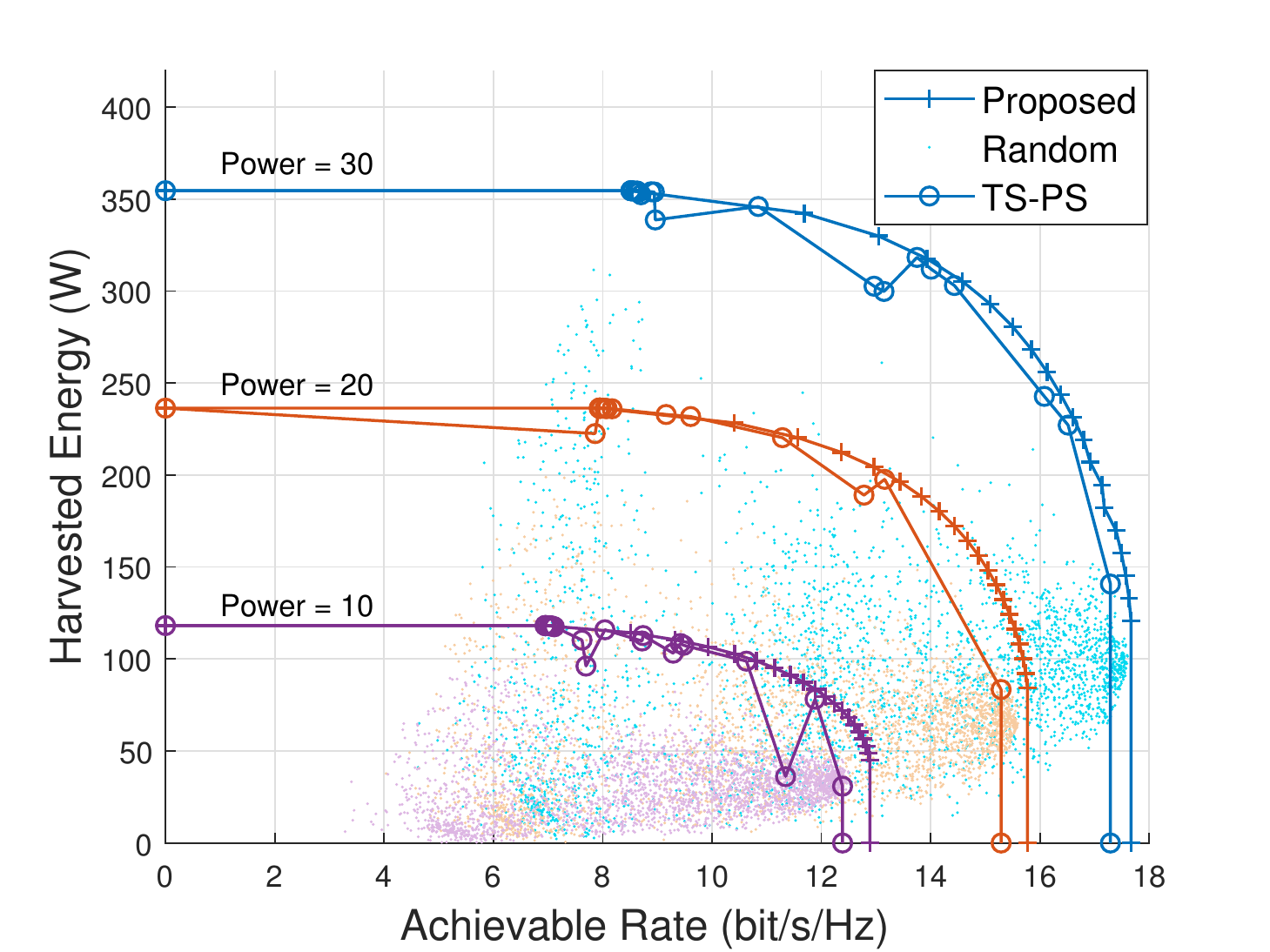}
		\label{fig_SWIPTcurve3}}
	\caption{Comparisons of SWIPT region among the 
		proposed  method, TS-PS and 10,000 random trials. 
		The color distinguishes  the transmit power $P$.} 
	\label{fig_SWIPTcomp}
\end{figure}

In Fig.~\ref{fig_SWIPTcomp}, the proposed CRM is compared with the
TS-PS \cite{zhang2013mimo} 
and random trials of $\mathbf{Q}$. For CRM and TS-PS,  eleven thresholds 
${\cal{\bar{E}}}$  equally dividing the interval $[{\cal E}_{\min},{\cal 
E}_{\max}]$ 
are considered in the cases of $P=10$, $20$, and $30$ (W).
In Fig.~\ref{fig_SWIPTcurve2}, 
the channels  are generated randomly as \eqref{eq_ch2}. The proposed CRM achieves the same performance as 
TS-PS. The random trials denoted as dots in Fig.~\ref{fig_SWIPTcurve2}  are based on 10,000  realizations of 
$\mathbf{Q}$.
In the case of $\mathbf{H}_1 = \mathbf{H}_2 = [1, 0.5; 0.5, 1]$, and 
$P=100$ (W), we found that the proposed method achieves the same 
rate-energy outer boundary and obtained the same figure with Fig. 7 in 
\cite{zhang2013mimo}. In Fig.~\ref{fig_SWIPTcurve3}, the channels are also randomly given as \eqref{eq_ch3}. Transmit through this channel, CRM can still reach the same rate-energy region, while TS-PS is not close to the upper rate-energy region. 
\begin{strip} \small
\begin{subequations}\label{eq_ch2}
	\begin{align} 
	&\mathbf{H}_1^{(a)} 
	= \left[\begin{matrix}
	0.45 + 1.75i& 0.85 - 1.26i&-0.52 + 0.33i&-0.83 + 0.30i\\
	-0.25 - 0.85i& 0.15 + 0.00i& -0.08 - 0.92i& 1.20 - 0.14i\\
	\end{matrix}\right],\\
	&\mathbf{H}_2^{(a)} 
	= \left[\begin{matrix}
	0.95 - 0.27i& -0.29 - 0.62i& 1.66 + 1.32i&-0.84 - 0.60i\\
	0.53 - 1.19i& 0.07 - 0.16i&-1.04 + 0.69i&-0.93 - 0.19i\\
	0.53 + 2.62i& 0.23 - 1.07i&-0.28 + 1.96i& 0.08 + 0.50i
	\end{matrix}\right].
	\end{align}
\end{subequations}
\begin{subequations}\label{eq_ch3}
	\begin{align}\small
	&\mathbf{H}_1^{(b)} 
	= \left[\begin{matrix}
	-2.13 - 1.49i& 1.85 + 0.80i&-0.96 + 1.51i& 1.55 - 0.28i\\
	0.00 - 1.43i&-1.16 + 0.19i&-0.27 + 0.53i& 0.10 + 0.12i\\
	-0.71 + 0.55i& 1.50 - 0.17i& 0.19 + 0.03i&-0.38 - 0.10i\\
	-0.08 - 1.44i& 0.80 + 0.13i&-0.45 + 1.51i& 0.25 + 0.98i
	\end{matrix}\right],\\
	&\mathbf{H}_2^{(b)} 
	= \left[\begin{matrix}
	-0.99 - 0.04i& 1. - 0.91i& 2.05 + 1.10i&-0.71 + 0.32i\\
	0.44 + 0.96i& 0.14 + 0.11i&-0.06 - 1.69i& 0.09 + 0.04i
	\end{matrix}\right].
	\end{align}
\end{subequations}
\end{strip}


The solution of SWIPT is a Pareto optimal among rate-energy pairs, however, TS-PS suffers from a non-convexity issue between the Lagrange multipliers and its objective function.  The performance of TS-PS relies on the initial selection of the Lagrange multipliers  and the algorithm may terminate at a local minimum. For TS-PS results presented in Fig.~\ref{fig_SWIPTcurve3}, we have chosen the best result (highest rate) after $20$ trials of different initial Lagrange multipliers. 
This shortcoming can not be ignored especially when  $m=4$ and $n_2\in\{1,2\}$. All these cases are practically important as the users usually have one or two antennas. In general, CRM is more robust and reliable for various numbers of antennas in the users. 

\begin{figure}[t]
	\centering
	\includegraphics[width=0.50\textwidth]
	{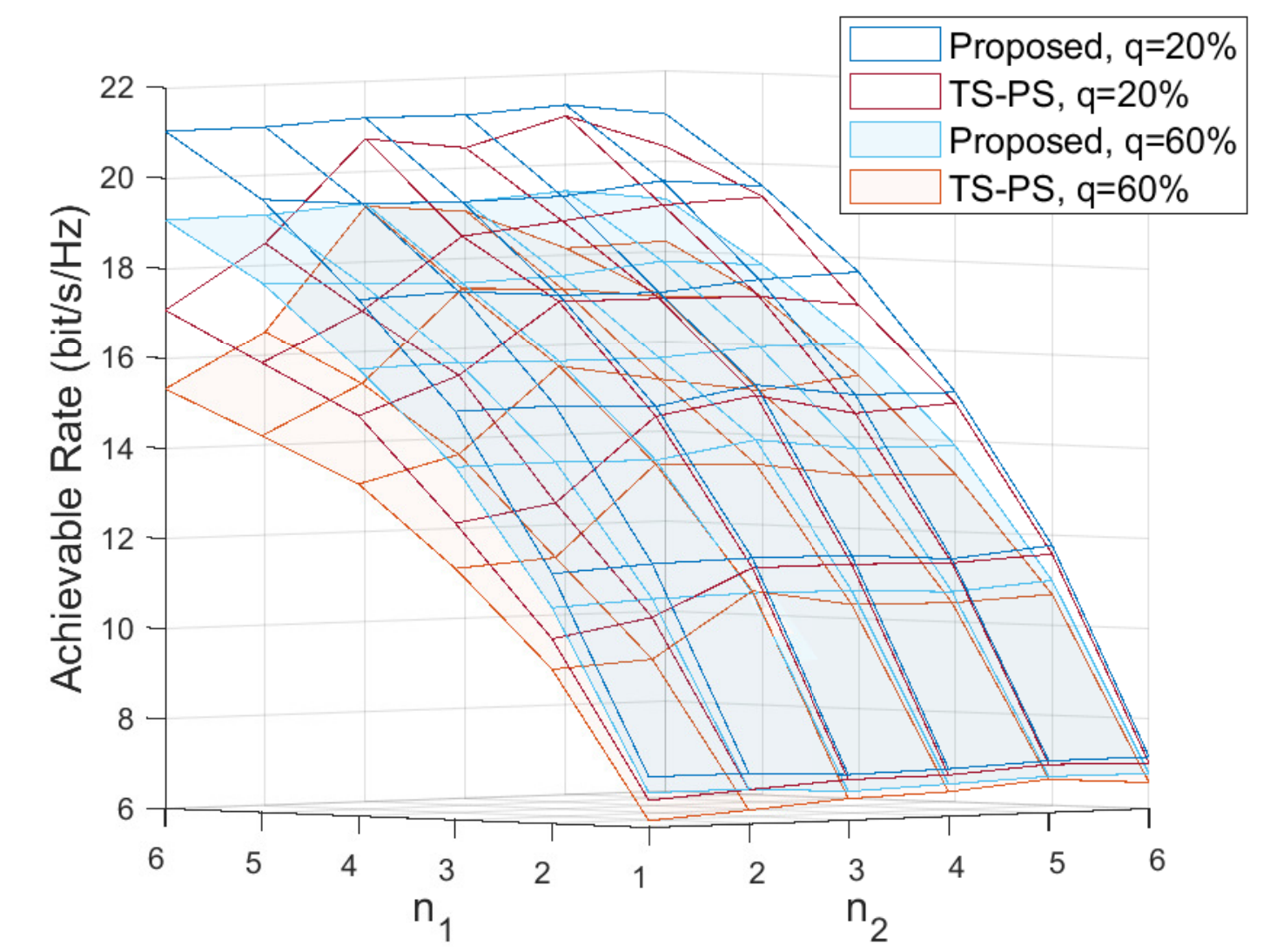}
	\caption{The achievable rate of CRM and TS-PS.}
	\label{fig_SWmap3}
\end{figure}

We compare the performance of CRM and  TS-PS under $m=4$, $P=20$ (W), and two thresholds, $q\in\{20\%, 60\%\}$, as representatives. The achievable rates are shown in Fig.~\ref{fig_SWmap3}. The two axes in the horizontal plane are $n_1$ and $n_2$; the vertical axis represents the rate achieved by Rx1. When $n_2\in\{1,2\}$, which refers to the left edges in Fig.~\ref{fig_SWmap3}, clear gaps can be seen between CRM and TS-PS for both $q=20\%$ and $q=40\%$. When $n_2$  raises, the gap shrinks, and the surfaces in Fig.~\ref{fig_SWmap3} almost overlap.

\begin{figure}[t]
	\centering
	\subfigure[Relative improvement ($\%$) when $q=20\%$.]{
		\includegraphics[width=0.46\textwidth]
		{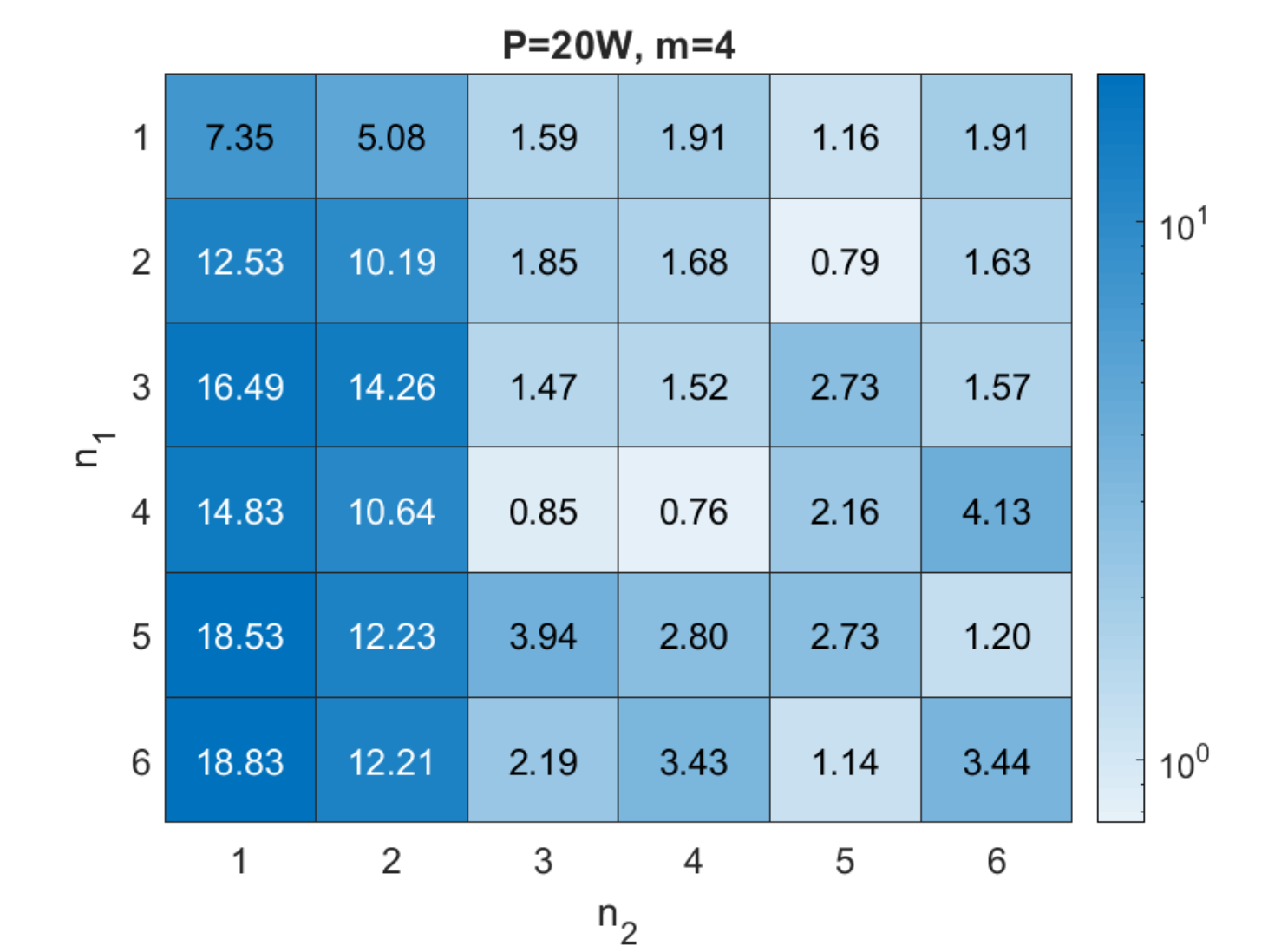}\label{fig_SWmap1}}
	\subfigure[Relative improvement ($\%$) when $q=60\%$.]{
		\includegraphics[width=0.46\textwidth]
		{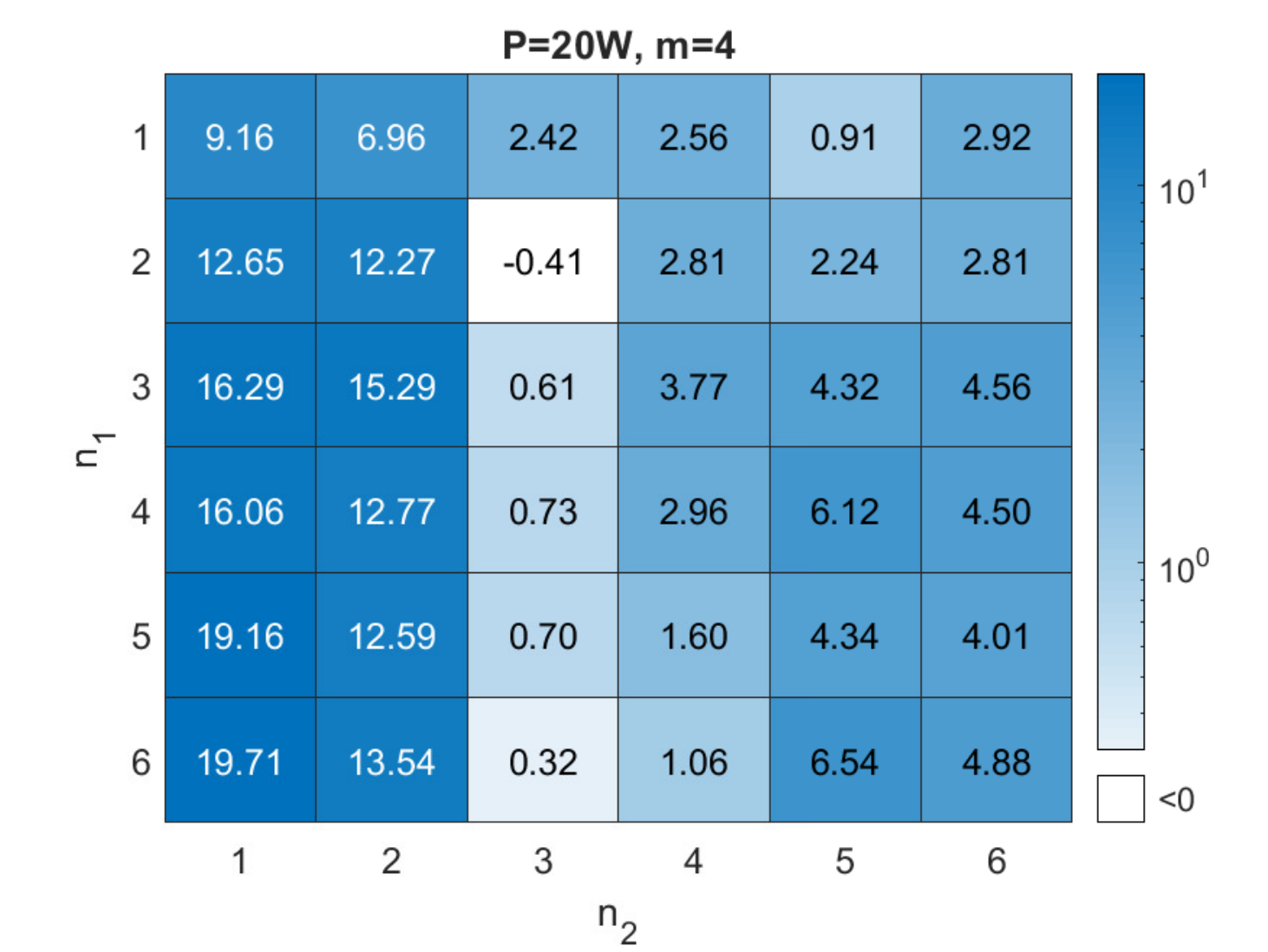}\label{fig_SWmap2}}
	\caption{Relative transmission rate improvement ($\%$) obtained by CRM  
		compared with  TS-PS solution for SWIPT for two different values of 
		$q$.}\label{fig_SWcomp}
\end{figure}

To make the comparison clearer, we quantitatively show the result by using \textit{relative improvement}, which is defined as
\begin{align}
	\eta_r=\frac{R_r-R}{R}\times100\%,\label{eq_diff_gsvd}
\end{align}
where $R_r$ and $R$ are the average rates achieved by 
CRM and the method compared with (for SWIPT, this method is TS-PS). 
A positive $\eta_r$ means an increased rate that CRM can achieve. 
The values in the heat maps refer to the.
We evaluate the relative improvement under various receiver antenna 
settings, more accurately $n_1,n_2\in\{1,\hdots,6\}$ and average over 100 channel 
realizations. Each entry of the channel matrices follows the normal 
distribution. A positive number means the proposed CRM reaches that much 
higher rate and 
in general, CRM is reliable in these cases. As can be seen from the first two columns of 
Fig.~\ref{fig_SWmap1}-\ref{fig_SWmap2}, the advantage of CRM is 
remarkable. These columns correspond to the cases in which the energy harvesting user has one or two antennas, which are the most common cases in practice.

\subsection{Multicasting}\label{sec_data_BC}
\begin{figure}[t]
	\centering
	\includegraphics[width=0.45\textwidth]
	{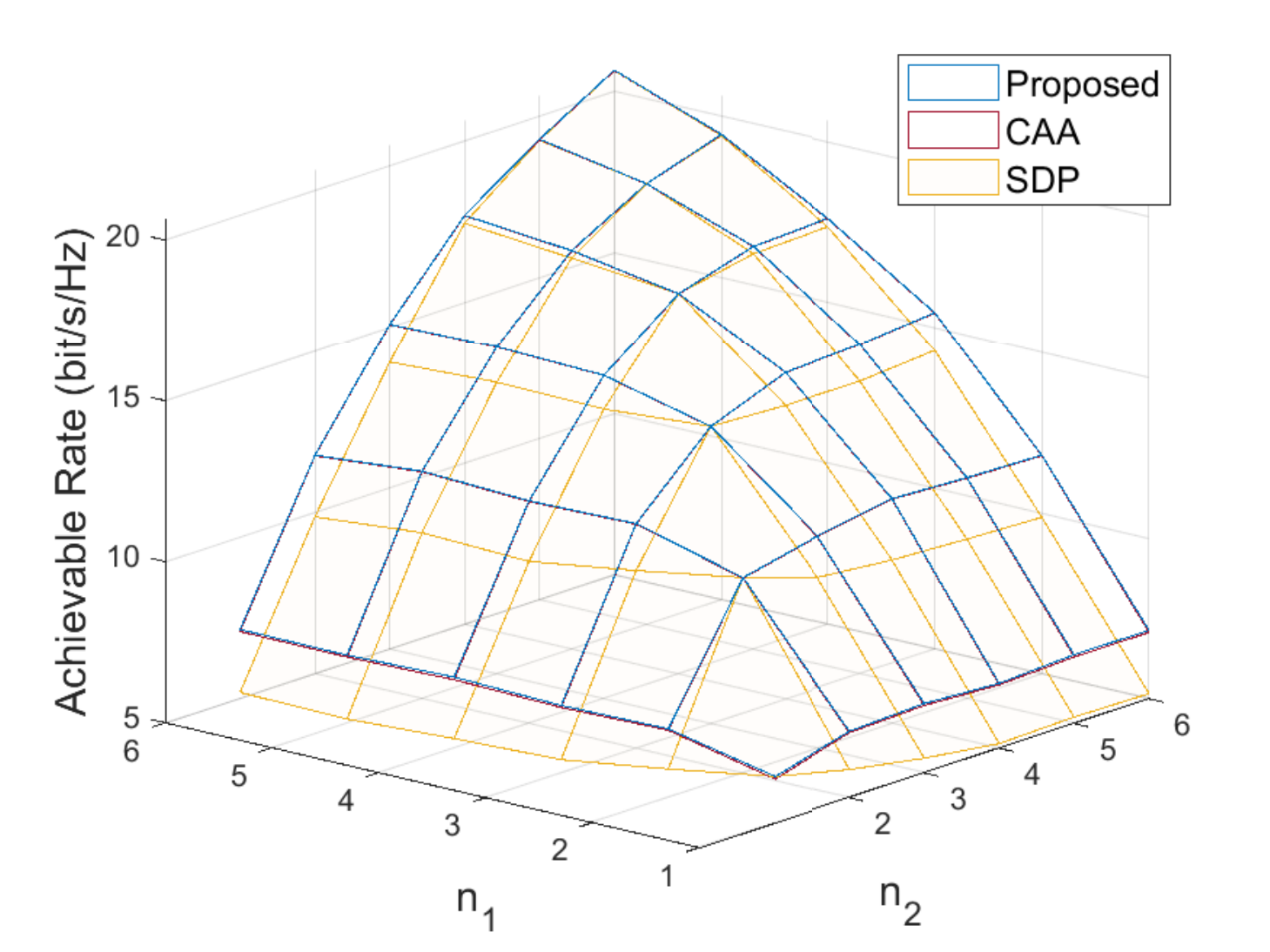}
	\caption{Achievable rates for CRM, CAA, and SDP.}
	\label{fig_BCmap3}
\end{figure}

\begin{figure}[t!]
	\centering
	\subfigure[Relative improvement ($\%$) of CRM to CAA.]{
		\includegraphics[width=0.41\textwidth]{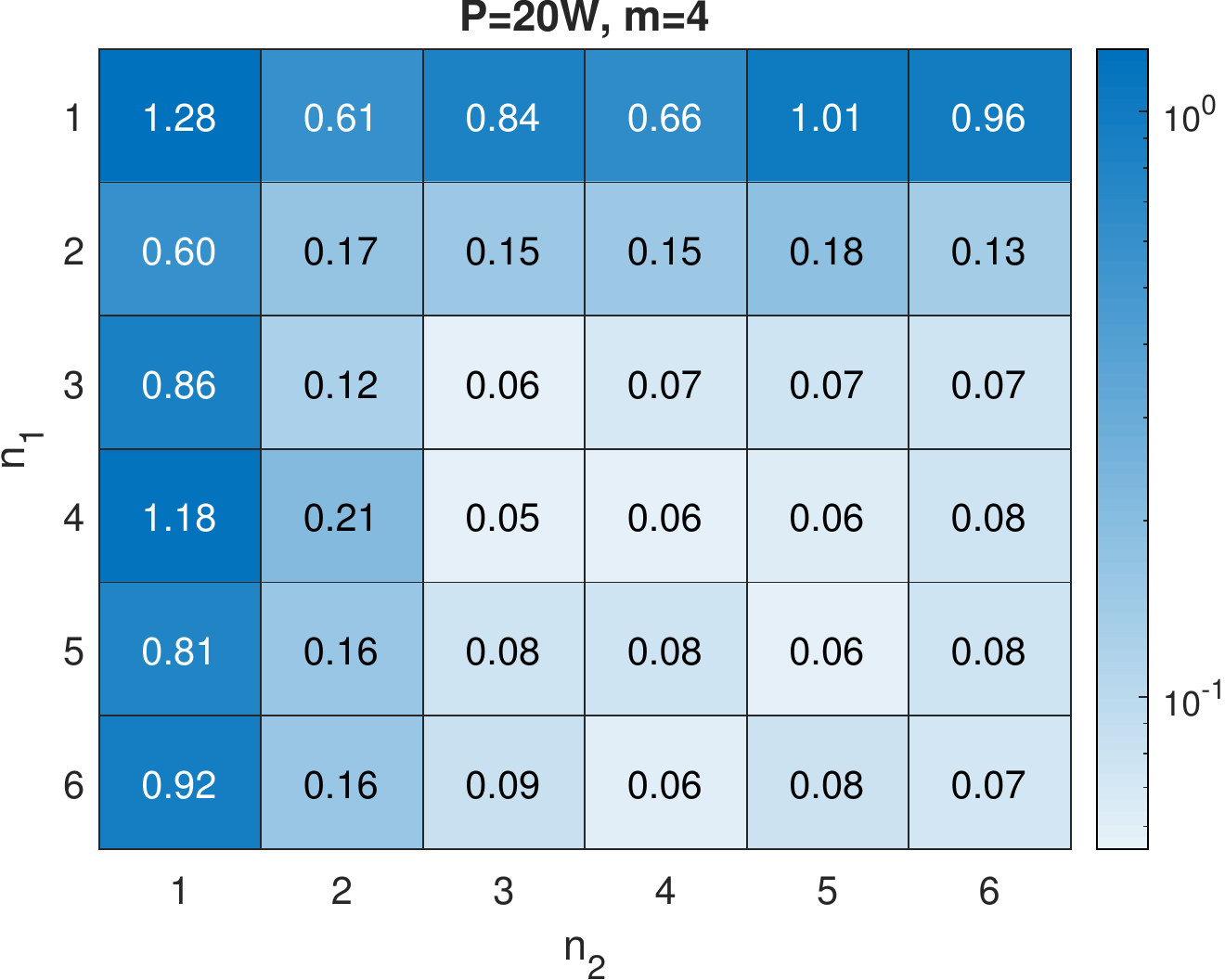}\label{fig_BCmap1}}
	\subfigure[Relative improvement ($\%$) of CRM to SDP.]{
		\includegraphics[width=0.46\textwidth]{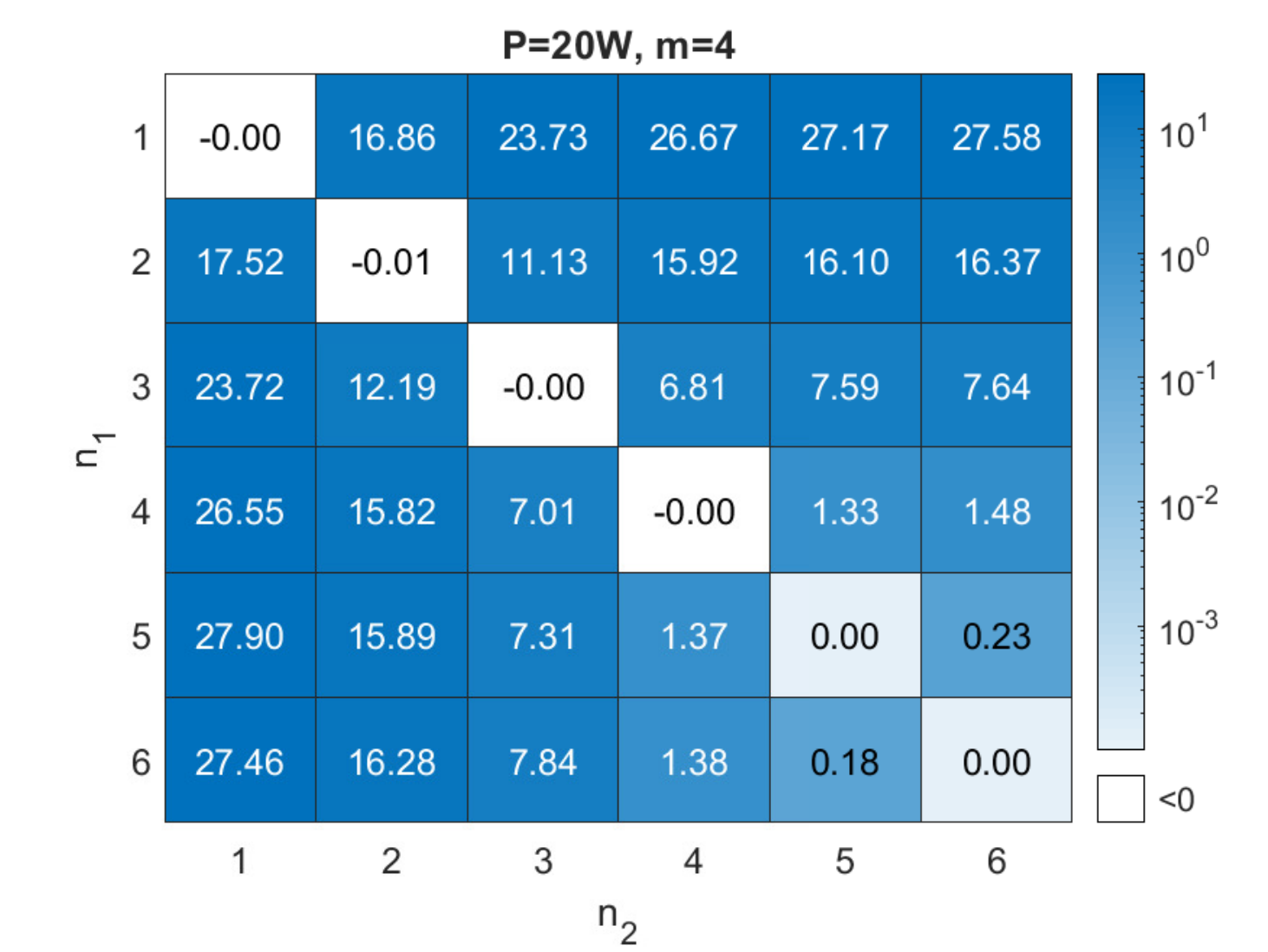}\label{fig_BCmap2}}
	\caption{Relative  multicasting  transmission rate improvement ($\%$)  
	obtained by CRM (the 
		proposed method) compared to the existing solutions (CAA and 
		SDP).}\label{fig_BCcomp}
\end{figure}

 To demonstrate the performance of CRM for multicasting, we 
	compare it with  CAA	 \cite{zhu2012precoder} and the  standard SDP 
	techniques solved by    \texttt{CVX} 	\cite{cvx}. 	
	We investigate a variety of combinations for $n_1, n_2\in\{1,\hdots,6\}$ 
	while $m=4$ and $P=20$ (W). 
	{The  multicasting rates achieved by those three methods are 
	shown in Fig.~\ref{fig_BCmap3}. The proposed method (CRM) can get the 
	same rate as 
	CAA. In comparison with SDP, clear gaps can be find when $n_1=1, 
	n_2\in\{2,\hdots,6\}$ or $n_2=1,n_2\in\{2,\hdots,6\}$.
	Again we use relative improvement defined in 
	\eqref{eq_diff_gsvd} and  the results are 
	shown in the heat maps in  Fig.~\ref{fig_BCcomp}.}
	It can be seen from Fig.~\ref{fig_BCmap1} that the proposed method gives
	a slight improvement to CAA. On the other hand, CRM outperforms SDP in most of the antenna settings  when  $n_1\neq n_2$, as shown in  Fig.~\ref{fig_BCmap2}. This improvement is especially remarkable when $n_1$ or $n_2$ is small.  
Such antenna settings are of great importance as real-world user 
equipment usually has one or two antennas.  

\begin{figure}[t!]
	\centering
	\includegraphics[width=0.475\textwidth]
	{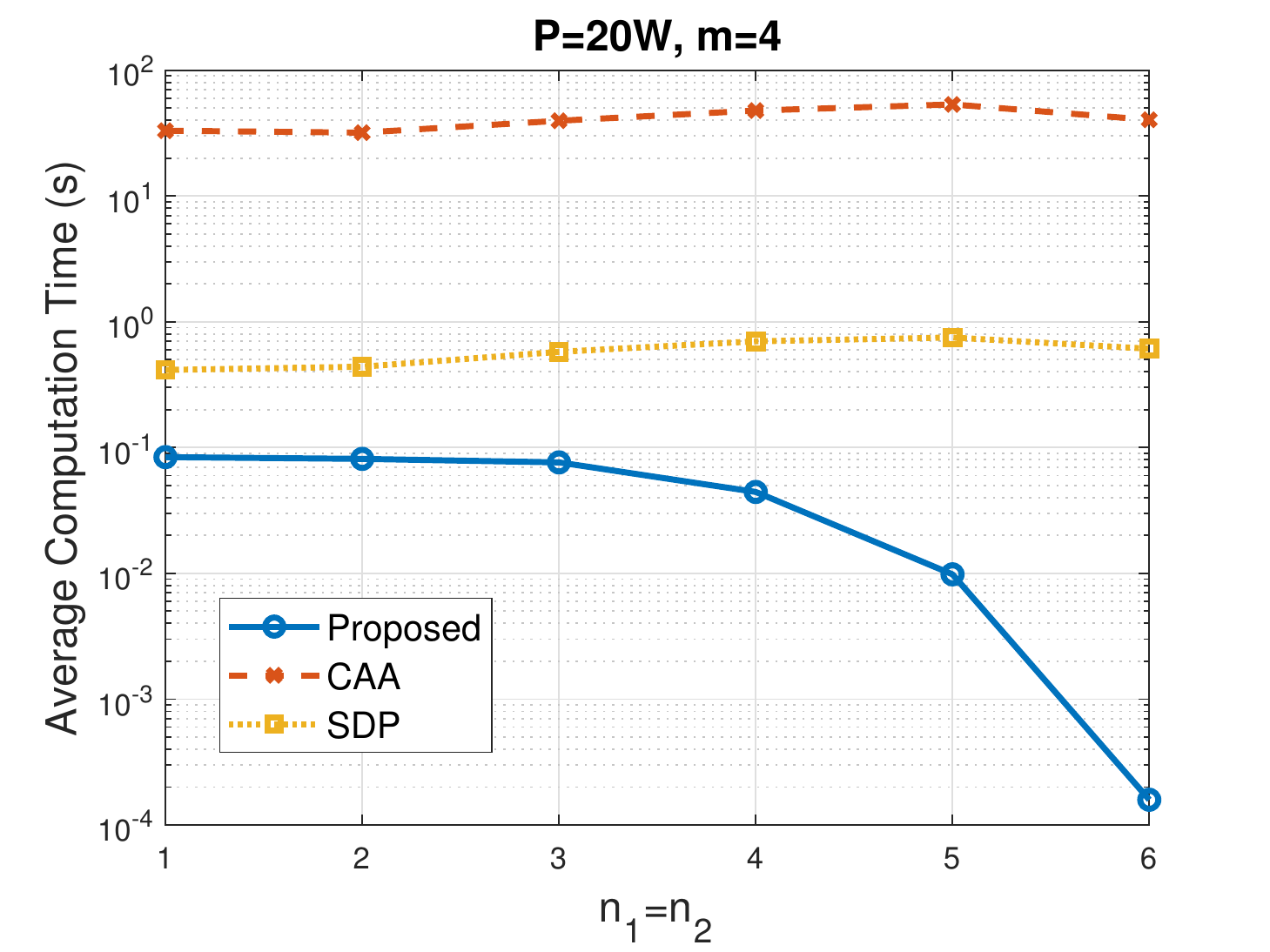}
	\caption{The computational cost of achieving multicasting.}
	\label{fig_BCcurve1}
\end{figure}

Besides, the merit of CRM in reducing the computational complexity should 
not  be ignored, which is 
analyzed in Fig.~\ref{fig_BCcurve1} by numerical simulations. 
This figure compares the time cost when $n_1=n_2$ where 
SDP works well. It can be seen that the proposed method has the best 
efficiency, while CAA is computationally expensive due to the successive 
optimization. The improvement of CRM with respect to CAA is more than 
three orders of magnitude which is incredibly high. Moreover,  
the time consumption of CRM goes down when $n_1=n_2$ becomes larger, whereas the time-cost of the other two methods slightly increases. 
{
Such an advantage is provided by  CRM for multicasting due to the three sub-cases  in Section~\ref{sec_rot_t7}. The probability of the first two sub-cases, which have an 
analytical solution, increases as $n_1=n_2$ becomes larger. As a result, the 
average time consumption of the proposed method reduces with the number 
of antennas.} Further, 
the results in Fig.~\ref{fig_BCmap3} indicate that the highest 
rate will be obtained when $n_1=n_2$, if $n_1+n_2$ is a constant. This is 
a big advantage indicating that CRM has both high rates  
and high time efficiency in practice.

In summary, the proposed rotation-based precoder provides a robust and 
efficient solution for both 	 problems studied in this paper, namely SWIPT and multicasting. The achievable rate is 
improved in both SWIPT and multicasting. The performance is no longer 
sensitive to the number of  antennas at the receiver. Meanwhile, the time 
efficiency of multicasting is significantly reduced.	

\section{Conclusion}\label{sec_conclu}
We have developed  complex rotation-based  precoding and power allocation for two different MIMO systems, namely,  SWIPT and physical layer multicasting. This method is  transforming positive semidefinite matrix constraints into a set of linear  
	constraints which is much simpler. The corresponding problem then can  be solved by  general optimization 
	methods rather than semidefinite programming solvers. For both problems,  the proposed 
	approach is feasible, reliable, and efficient  and 
	outperforms state-of-the-art solutions noticeably. 
	
The complex rotation-based precoding approach developed in this paper can be applied to  other emerging problems such as  SWIPT with multi-group model \cite{gautam2019multigroup} and  unmanned aerial 
		vehicle and non-orthogonal multiple access scenario \cite{wang2020joint}, and 
	  MIMO systems with finite alphabet 
			 \cite{xiao2011globally},  
		 to name  a few.

	\balance
	\typeout{}
	\bibliography{REF_commu_v1.0_nolink}
	\bibliographystyle{ieeetr}

\end{document}